\documentclass{article}
\usepackage{a4wide,amssymb,amsmath,cite}
\title{Gluon Correlators in the Kogan--Kovner Model}
\author{B. M. Gripaios \thanks{b.gripaios1@physics.ox.ac.uk}
\\\emph{Department of Physics - Theoretical Physics, University of Oxford,} \\ \emph{1 Keble Road, Oxford. OX1 3NP  UK}
\\
\\ OUTP-02 22P}
\date{April 26, 2002}
\begin{document}
\bibliographystyle{h-elsevier2}
\maketitle

\begin{abstract}
The Lorentz-invariant gluon correlation functions $\langle F^2(x) F^2(x') \rangle$
and $\langle F \tilde{F} (x) F \tilde{F} (x') \rangle$, corresponding to scalar and pseudo-scalar glueballs,
are calculated for Kogan's and Kovner's variational ansatz for the pure $SU(N)$ Yang--Mills wavefunctional.

One expects that only one dynamical mass scale should be present in QCD;
the ansatz generates the expected scale for both glueballs, as well as an additional scale for the scalar glueball.
The additional mass scale must therefore vanish,
or be close to the expected one. This is shown to constrain the nature of the phase transition in the
Kogan--Kovner ansatz.
\end{abstract}

\section{Introduction}
Strongly interacting non-Abelian gauge theories require non-perturbative methods.
One such, the application of the variational approximation to the Schr\"{o}dinger formulation,
goes back to the days of Feynman \cite{Feynman:1987hv}. Though such methods are approximate, 
they incorporate many of the desirable features of non-Abelian gauge theories, 
in particular the gluon condensate, asymptotic freedom, instantons, dynamical mass generation and confinement 
\cite{Kogan:1995wf, Brown:1997gm, Zarembo:1998ms, Brown:1998nz, Diakonov:1998ir, Brown:1998cp}. 
Variational methods have also met success applied to Abelian theories \cite{Drell:1979hr, Kogan:1995vb, Kovner:1998eg}.

In the Schr\"{o}dinger formulation of pure $SU(N)$ Yang--Mills theory, the wavefunctional 
$\Psi [ A^{a}_{i} (x) ]$ 
is an eigenstate of the Hamiltonian
\begin{gather} 
H = \int d^3 x \; \frac{1}{2} [E^{a 2}_{i} (x) + B^{a 2}_{i} (x)], 
\end{gather}
where 
$E^{a}_{i} (x) = i \frac{\delta}{\delta A^{a}_{i} (x)} $, 
$B^{a}_{i} (x) = \frac{1}{2} \epsilon_{ijk} (\partial_j A^{a}_{k} (x) -  
\partial_k A^{a}_{j} (x) + g f^{abc} A^{b}_{j} A^{c}_{k} ) $
and physical states satisfy Gauss' law as a constraint:
\begin{gather}
[ \partial_i E^{a}_{i} (x) - g f^{abc} A^{b}_{i} E^{c}_{i} ] \Psi [ A ] = 0.
\end{gather}
Kogan and Kovner \cite{Kogan:1995wf} proposed the gauge invariant Gaussian ansatz
\begin{gather}
\Psi [ A^{a}_{i} (x) ] = \int D U \;
\exp \Big[ - \frac{1}{2} \int d^3 x d^3 y \; A^{U a}_{i} (x) \delta^{ab} \delta_{ij} G^{-1} (x-y) A^{U b}_{j} (y) \Big],
\end{gather}
where under a local gauge transformation $U(x),$
\begin{gather}
A^{a}_{i} (x) \rightarrow A^{U a}_{i} (x) = S^{ab} (x) A^{b}_{i} (x) + \lambda^{a}_{i} (x),
\end{gather}
and $D U$ is the $SU(N)$ group invariant measure.
Here $S^{ab} = \frac{1}{2} \mathrm{tr} ( \tau^a U^{\dag} \tau^b U )$, 
$\lambda^{a}_{i} = \frac{i}{g} \mathrm{tr} ( \tau^a U^{\dag} \partial_i  U )$ 
and 
$\frac{\tau^a}{2}$ form an $N \times N$ Hermitian representation of $SU(N)$: 
$[ \frac{\tau^a}{2} , \frac{\tau^b}{2} ] = i f^{abc} \frac{\tau^c}{2}$ with normalisation 
$\mathrm{tr} ( \tau^a \tau^b ) = 2 \delta^{ab}.$
 The $A$-field propagator has a mass parameter $M$, with respect to which the Hamiltonian is minimised.
 The authors found that $M$ is of the correct size to generate the observed gluon condensate \cite{Shifman:1979bx}.
 
In sections \ref{sec:scalar} and \ref{sec:pseudo} of this paper, the Lorentz invariant gluon correlators 
$\langle F^2(x) F^2(x') \rangle$ and $\langle F \tilde{F} (x) F \tilde{F} (x') \rangle$
are calculated using the same ansatz, and their
scaling properties at large separations are analysed. Na\"{\i}vely, one expects that the correlators will decay
as $\exp - M |x - x'|$, with mass scale $M$. 
Firstly though, let us recall the details of the ansatz and calculation of the correlation functions therein.
\section{Correlator Rubric}
To calculate correlation functions, consider the generating functional
\begin{gather}
Z = \int D A \; \Psi [ A ]  \Psi [ A ];
\end{gather}
the correlator of operator $O$ is then
\begin{gather} \label{corrdef}
\langle O \rangle = \frac{1}{Z} \int D A \; \Psi [ A ] O \Psi [ A ].
\end{gather}
For gauge invariant $O$, one of the gauge integrations is redundant. Since the gauge transformation is linear on the gauge
fields, the gauge field path integral is Gaussian. Using matrix notation for integration, \emph{viz.}
\begin{gather}
(AB)^{ac}_{ik} (x,z) = \Sigma_{b,j}\int d^3 y \; A^{ab}_{ij} (x,y) B^{bc}_{jk} (y,z),
\end{gather}
the generating functional is proportional to
\begin{gather} \label{genfunc}
\int D A D U \; \exp [- \frac{1}{2} (A+a)^T \mathcal{M} (A+a)] 
\exp  [ - \frac{1}{2} \lambda^T (G + SGS^T)^{-1}) \lambda ].
\end{gather}
Here, matrices are defined as
\begin{align} \label{Mprop}
S^{ab}_{ij}(x,y) &= S^{ab} (x) \delta (x-y) \delta_{ij}, \nonumber \\ 
G^{ab}_{ij} &= G (x-y) \delta^{ab} \delta_{ij}, \nonumber \\ 
\mathcal{M} &= S^T G^{-1} S + G^{-1} \; \mathrm{and} \nonumber \\
a &= \mathcal{M}^{-1} S^T G^{-1} \lambda.
\end{align}
Note that $S$ is orthogonal whereas $\mathcal{M}$ and $G$ are symmetric.

To calculate a correlator, substitute the operator in (\ref{corrdef})
to obtain the correlator of some string of $A$ fields. 
Then do the $A$-field path integral; this is a free 
field theory. Thus
\begin{gather} \label{Aprop}
\langle A ^{a}_{i} (x) A ^{b}_{j} (y) \rangle = a^{a}_{i} (x) a^{b}_{j} (y) + \mathcal{M}^{-1ab}_{\phantom{-1}ij} (x,y),
\end{gather}
and so on via Wick's theorem.

This done, one is left with an integral over gauge transformations $U$. The stratagem here is to 
factor the gauge transformation $U$ into a product $U_L U_H$ with parts dependent on high or low 
momentum modes of magnitude $k$: $U_H$ with $k > M$ and $U_L$ with $k < M$ respectively. 
Both the measure and action factorise similarly, since the group measure is invariant 
and the high and low modes are orthogonal. 
The $A$-field propagator takes the form $G^{-1} (k) = \sqrt{ k^2 + M^2}$.
For $k > M$, take 
$G^{-1} (k) \simeq k$ and treat the theory perturbatively. With $U = \exp [i g \phi^a \tau^a] $,
to $O (g ^2) $ the high mode action takes the form
\begin{gather}  \label{himode}
\Gamma_H [\phi]= \frac{1}{4} \int d^3x d^3y \; \partial_i \phi^a (x) \left. G^{-1} (x-y) \right|_{k>M} \partial_i \phi^a (y).
\end{gather}
One again obtains a free field, with propagator
\begin{gather} \label{hiprop}
\langle \phi^a (x) \phi^b (y) \rangle = 2 \delta^{ab}  [ \left. \partial^{x}_{i} \partial^{y}_{i} G^{-1} (x-y) \right|_{k>M}]^{-1} 
\equiv 2 \delta^{ab} H^{-1} (x-y) .
\end{gather}
For $k < M$, take $G^{-1} (k) \simeq M$  such that $G^{-1} (x-y)|_{k<M}$ is local;
the matrices $U_L$ vary only on scales greater than $1/M$ 
and can be considered constant in the action.\footnote{In the action, $U_L$ are multiplied by $G^{-1} (x) \sim 0$ for $|x|>1/M$} 
The resulting leading order low mode action is  
\begin{gather} \label{lomode}
\Gamma_L[U] = \frac{M}{2 g^2} \mathrm{tr} \int d^3x \; \partial_i U^{\dag}_{L} (x) \partial_i U_{L} (x).
\end{gather}
The integration over high modes amounts to a renormalization group transformation. 
The effect on the low modes is to replace $g$ in (\ref{lomode}) by the coupling at the scale $M$.
 This sigma model is calculated in the mean field approximation: the action is replaced by
\begin{gather} \label{lomode2}
\frac{M}{2 g^2(M)} \mathrm{tr} \int d^3x \; \big[\partial_i U^{\dag}_{L} (x) \partial_i U_{L} (x) + 
\sigma^2 (U^{\dag}_{L} U_{L} - 1)\big],
\end{gather}
subject to the constraints
\begin{gather}
\langle U^{\dag}_{L} U_{L} \rangle = 1,   \\
\langle \sigma U_{L} \rangle = 0.
\end{gather} 
So the $U_{mn} (x)$ act as $N^2$ free charged scalar fields with propagator
\begin{gather} \label{loprop}
\langle U^{\dag}_{Lkl} (x) U_{Lmn} (y) \rangle = \frac{2 g^2}{M}\delta_{kn}\delta_{lm}   (-\partial_i \partial_i + \sigma^2 )^{-1} \delta (x-y).
\end{gather}

The momentum split allows the further simplifications
\begin{align} \label{approx}
S^{ab} (x) &= S^{ab}_{L} (x), \nonumber\\
\mathcal{M}^{ab} (x,y) &= 2 \delta^{ab} G^{-1} (x-y),   \nonumber\\
\lambda^{a}_{i} (x) &= \lambda^{a}_{iL} (x) + \lambda^{a}_{iH} (x), \nonumber\\
a^{a}_{i} (x) &= \frac{1}{2} \lambda^{a}_{iL} (x) + \frac{1}{2} \lambda^{b}_{iH} S^{ba}_{L} (x).
\end{align}

Before tackling specific correlators, let us note also the following points:
\begin{itemize}
\item The non-perturbative physics is captured in the low mode action; elsewhere we work to $O(g)$.
\item A factor $S_L$ in a correlator contributes $\langle U^{\dag}_{L} U_{L} \rangle$
in the Wick expansion, and therefore a factor of $g^2$,
from (\ref{loprop}). The only leading order terms are those in which the orthogonality of the $S_L$ can be used, such that they cancel.
\item Orthogonality of the $S_L$ can be used whenever they are separated by a scale less than $1/M$. Thus, wherever, $S_L$ appears, 
it is sufficient to assume orthogonality, and integrate up to distance scale $1/M$. Contributions on larger scales are higher order in $g$. 
\item The action is quadratic in the decoupled $\lambda_L,\lambda_H$; correlators with an odd number of $\lambda_L$ or $\lambda_H$ vanish.
\item A correlator with an odd number of $a$ and \emph{ergo} $A$ will therefore vanish to leading order.
\item Since $\lambda^{a}_{iH}$ is just $-\partial_i \phi^{a}$, $\epsilon_{ijk} \partial_j \lambda^{a}_{kH} = 0$.
\item $\lambda_L$ is pure gauge and Abelian magnetic terms like $(\epsilon_{ijk} \partial_j \lambda^{a}_{kL})^2$ vanish to $O(g)$.
\item Thus, any correlator with two or more $\epsilon\partial a$ terms must vanish to $O(g)$. 
\item In seeking the scaling properties of correlators, constants and contact terms may be discarded.
\end{itemize}

\section{Scalar gluon correlator} \label{sec:scalar}
The Lorentz invariant gluon correlators are $\langle F^2(x) F^2(x') \rangle$
and $\langle F \tilde{F} (x) F \tilde{F} (x') \rangle$, where
\begin{gather}
F^2 = F^{a}_{\mu\nu} F^{a\mu\nu} = 2 ( B^{a 2}_{i} - E^{a 2}_{i} ), \\
F \tilde{F} = \frac{1}{2!}\epsilon_{\mu\nu\sigma\rho} F^{a}_{\mu\nu} F^{a}_{\sigma\rho} = 4 B^{a}_{i} E^{a}_{i}.
\end{gather}
Discarding a contact term, the scalar correlator is thus
\begin{gather}
\langle F^2(x) F^2(x') \rangle = 4 \langle B^2(x) B^2(x') + E^2(x)E^2(x') - 2 B^2(x) E^2(x')\rangle.
\end{gather}
To $O(g)$, the chromo-magnetic term is
\begin{align}
B^{a 2}_{i} (x) B^{'a 2}_{i} (x) &= \epsilon_{ijk}\epsilon_{ilm}\epsilon^{'}_{ijk}\epsilon{'}_{ilm}
[\partial_j A^{a}_{k} \partial_l A^{a}_{m} \partial^{'}_{j} A^{'a}_{k} \partial^{'}_{l} A^{'a}_{m} \nonumber \\
&+ g f^{abc} \partial_j A^{a}_{k} A^{b}_{l} A^{c}_{m} \partial^{'}_{j} A^{'a}_{k} \partial^{'}_{l} A^{'a}_{m}
+ g f^{'abc} \partial{'}_{j} A^{'a}_{k} A^{'b}_{l} A^{'c}_{m} \partial_j A^{a}_{k} \partial_l A^{a}_{m}],
\end{align}
where a prime on a field indicates that its parameters carry primes.
The terms of $O(g)$ have five $A$-fields and these vanish. Using (\ref{Aprop}), the only terms with less than two
$\epsilon\partial a$s contain $\mathcal{M}$s only. To wit,
\begin{align}
\langle B^{a 2}_{i} (x) B^{'a 2}_{i} (x) \rangle = \epsilon_{ijk}\epsilon_{ilm}\epsilon^{'}_{ijk}\epsilon^{'}_{ilm}
[&\partial_{j}^{x}\partial_{l}^{y}\mathcal{M}^{-1aa}_{\phantom{-1}km} (x,y) 
\partial_{j}^{'x}\partial_{l}^{'y}\mathcal{M}^{'-1aa}_{\phantom{-1}km} (x,y)|_{x=y} \nonumber \\
+&\partial_{j}^{x}\partial_{j}^{'x}\mathcal{M}^{-1aa'}_{\phantom{-1}kk'} (x,x')
 \partial_{l}^{x}\partial_{l}^{'x}\mathcal{M}^{-1aa'}_{\phantom{-1}mm'} (x,x')      \nonumber \\
+&\partial_{j}^{x}\partial_{l}^{'x}\mathcal{M}^{-1aa'}_{\phantom{-1}km'} (x,x') 
 \partial_{l}^{x}\partial_{j}^{'x}\mathcal{M}^{-1aa'}_{\phantom{-1}mk'} (x,x')].
\end{align}
Using $\epsilon_{ijk}\epsilon_{ilm} = \delta_{jl}\delta_{km} - \delta_{jm}\delta_{kl}$ and
$\mathcal{M}^{ab}_{ij}(x,y) = 2 \delta^{ab} \delta_{ij} G^{-1} (x-y)$ yields (restoring primes and discarding a constant)
\begin{align} \label{bbbb}
+\frac{1}{2}&(N^2-1)\partial_{j}^{x}\partial_{j'}^{x'} G(x-x') \partial_{j}^{x}\partial_{j'}^{x'} G(x-x') \nonumber \\
+\frac{1}{2}&(N^2-1)\partial_{j}^{x}\partial_{j}^{x'} G(x-x') \partial_{j'}^{x}\partial_{j'}^{x'} G(x-x').
\end{align}

How do these terms scale? Recall
\begin{gather}
G(x) = \int^{\infty}_{-\infty} \frac{d^3 k}{(2\pi)^3}\; G(k) e^{ik.x},
\end{gather}
where $G(k)^{-1} = \sqrt{k^2+M^2}$. Choose spherical polars with $k_3$-axis along $x$. Then
\begin{gather}
G(x) = \frac{1}{2\pi^2x} \int_{0}^{\infty} dk \; \frac{k}{\sqrt{k^2+M^2}} \sin kx = 
 \frac{1}{4\pi^2ix} \int_{-\infty}^{\infty} dk \; \frac{k}{\sqrt{k^2+M^2}} e^{ikx}.
\end{gather}
The contour can be deformed in $\mathbb{C}$ to surround the branch cut going
 from $iM$ to $\infty$ (\emph{cf.} \cite[pp. 27-28]{Peskin:1995ev}) and for large $x$ approaches
\begin{gather}
\frac{1}{2\pi^2x} \int_{M}^{\infty} dy \frac{ye^{-xy}}{\sqrt{y^2-M^2}} = \frac{M}{2\pi^2x}K_1(Mx)
 \sim (\frac{M}{8\pi^3x^3})^{1/2} e^{-Mx} .
\end{gather}
Similarly,
\begin{gather}
G^{-1}(x) = \frac{1}{4\pi^2ix} \int_{-\infty}^{\infty} dk \; k\sqrt{k^2+M^2} e^{ikx}
= \frac{M^2}{2\pi^2x^2}K_2(Mx) 
\sim (\frac{M^3}{8 \pi^3 x^5})^{1/2} e^{-Mx},
\end{gather}
where $K_{1,2}(Mx)$ are modified Bessel functions \cite{Watson}. 

Carrying out the differentiation in (\ref{bbbb}), one finds that
\begin{gather} \label{chromomag}
4 \langle B^2(x)B^2(x') \rangle \sim \frac{4(N^2-1) M^5 e^{-2M|x-x'|}}{(2\pi)^3 |x-x'|^3},
\end{gather}
where the expected mass scaling is indeed present.

Consider now the chromo-electric correlator. Here things become more complicated. Doing the functional differentiation 
in (\ref{corrdef}), one obtains
\begin{align}
\langle E^{a 2}_{i} (x) E^{'a 2}_{i} (x) \rangle &= 6 (N^2-1) [G^{-1}(x-x')]^2 \nonumber \\
&- 4 G^{-1}(x-x') \int d^3y d^3y' \; G^{-1}(y'-x')G^{-1}(y-x) \langle A^{a}_{i} (y) A^{a}_{i} (y')\rangle \nonumber \\
&+ \int d^3y d^3y' d^3z d^3z' \; G^{-1}(y'-x') G^{-1}(z'-x') G^{-1}(y-x) G^{-1}(z-x) \nonumber \\
&\langle A^{a'}_{i'} (y') A^{a'}_{i'} (z') A^{a}_{i} (y) A^{a}_{i} (z) \rangle,
\end{align}
plus constants. Using (\ref{Aprop}) and (\ref{approx}), and the orthogonality properties of the $S_L$, this becomes
\begin{align} \label{ellam}
&\frac{3}{2} (N^2-1) [G^{-1}(x-x')]^2 \nonumber \\
- &\frac{1}{2} G^{-1}(x-x') \int d^3y d^3y' \; G^{-1}(y'-x')G^{-1}(y-x) 
[ \lambda^{a}_{iL} (y') \lambda^{a}_{iL} (y) + \lambda^{a}_{iH} (y') \lambda^{a}_{iH} (y)] \nonumber \\
+& \frac{1}{16}\int d^3y d^3y' d^3z d^3z' \; G^{-1}(y'-x') G^{-1}(z'-x') G^{-1}(y-x) G^{-1}(z-x) \nonumber \\
&[\lambda^{'a}_{iL} (y) \lambda^{'a}_{iL} (z) \lambda^{a}_{iL} (y) \lambda^{a}_{iL} (z)
+\lambda^{'a}_{iH} (y) \lambda^{'a}_{iH} (z) \lambda^{a}_{iH} (y) \lambda^{a}_{iH} (z)
+4\lambda^{'b}_{iH} (y) S^{'ba}_{iL} (y)\lambda^{'a}_{iL} (z) \lambda^{b}_{iH} (y) S^{ba}_{iL} (y) \lambda^{a}_{iL} (z)].
\end{align}
The inverse relation 
$\int d^3y \; G^{-1}(x-y) G (y-z) = \delta^3 (x-z)$ has been used.
 
The averaging over the gauge transformations $U$ remains. The orthogonality of the
 $k$-modes ensures the following property: in any integral of the form $\int d^3x \; f(x) g(x)$, where 
 $g$ contains only high (low) mode contributions, only high (low) mode parts of $f$
 contribute to the integral. Then, since $G^{-1}(x-y)|_{k<M} \simeq M \delta (x-y)$ is local, the integrations in
 (\ref{ellam}) involving $\lambda_L$ are trivial. 
 
The integrations involving $\lambda_H$ are a little more tricky. Consider, for example, the simplest one:
\begin{gather}
-\frac{1}{2}G^{-1}(x-x')\int d^3y d^3y' \; G^{-1}(y'-x')G^{-1}(y-x) \langle \lambda^{a}_{iH} (y') \lambda^{a}_{iH} (y) \rangle.
\end{gather}
Using (\ref{hiprop}), this is
\begin{gather}
- (N^2-1) G^{-1}(x-x')\int d^3y d^3y' \; G^{-1}(y'-x')G^{-1}(y-x) \partial^{y}_{i} \partial^{y'}_{i}H^{-1}(y-y').
\end{gather}
Fourier decomposition of the propagators yields the simple result
\begin{gather} \label{hiquad}
- (N^2-1) G^{-1}(x-x') G^{-1} (x-x')|_{k>M}.
\end{gather}
The quartic term is similar. Since 
\begin{align}
&\langle \phi^{'a} (y) \phi^{'a} (z) \phi^{a} (y) \phi^{a} (z) \rangle =
4 \big[(N^2-1)^2 H^{-1}(y'-z') H^{-1}(y-z) \nonumber \\
&+ (N^2-1) H^{-1}(y'-y) H^{-1}(z'-z)
+ (N^2-1) H^{-1}(y'-z) H^{-1}(z'-y)\big],
\end{align}
the terms dependent on $x-x'$ become (Fourier decomposing again)
\begin{gather} \label{hiquar}
\frac{(N^2-1)}{4} \int_{k>M} \frac{d^3k}{(2\pi)^3} \frac{d^3k'}{(2\pi)^3} \;
\frac{G^{-1}(k) G^{-1}(k')}{k^2 k^{'2}}(k.k')^2 e^{i(k+k').(x - x')}.
\end{gather}
These integrals can be evaluated, but are not very illuminating, amounting to Fourier transforms of various simple powers of $k$. 
If the theory is regularised with a cut-off, these are just spurious oscillatory functions of $M|x-x'|$.
 One obtains the same generic oscillatory result if one tries to calculate $G^{-1} (x-x')|_{k\lessgtr M}$ explicitly;
 what is important for our purposes is that the only mass scale present is $M$.

The remaining term in $\lambda_H$ in (\ref{ellam}) is quadratic. The high mode propagator contributes a factor $\delta^{bb'}$
 and the orthogonality property of the $S_L$ can then be used.
 
The low mode expectation values remain in (\ref{ellam}). The simplest terms are quadratic:
\begin{gather}
\langle\lambda^{a}_{iL} (x') \lambda^{a}_{iL} (x)\rangle.
\end{gather}
The Fierz identity for $SU(N)$:
$\tau^{a}_{ij}\tau^{a}_{kl} = 2 (\delta_{il}\delta_{jk}-\frac{1}{N}\delta_{ij}\delta_{kl})$ implies
\begin{gather}
\lambda^{a}_{iL} (x')\lambda^{a}_{iL} (x) =
-\frac{2}{g^2}\mathrm{tr} (U^{\dag}_{L} (x')\partial_i U_L (x')U^{\dag}_{L} (x) \partial_i U_L (x)).
\end{gather}
The $U_L$ contain modes with $k<M$; for $x\neq x'$, the unitarity of the $U_L$ cannot be invoked. 
Doing the Wick expansion, one sees that these term are $O(g^2)$.

The quartic low mode term is (using (\ref{loprop}))
\begin{align} \label{loquar}
&\frac{M^4}{16}\langle\lambda^{'a}_{iL} (x) \lambda^{'a}_{iL} (x) \lambda^{a}_{iL} (x) \lambda^{a}_{iL} (x)\rangle \nonumber \\
=&\frac{M^4}{4g^4}\langle \mathrm{tr}(\partial^{'}_{i} U (x') \partial^{'}_{i} U^{\dag} (x'))
\mathrm{tr}(\partial_{i} U (x) \partial_{i} U^{\dag} (x))\rangle \nonumber \\
= &2M^2N^2 \left[ \int_{k<M} \frac{d^3k}{(2\pi)^3} \; \frac{k_ik_{i'}}{k^2+\sigma^2} e^{ik.(x-x')} \right]^2.
\end{align}

This low mode term is very interesting. Consider the integral
\begin{gather}
\int_{k<M} \frac{d^3k}{(2\pi)^3} \; \frac{e^{ik.x}}{k^2 + \sigma^2}=
\frac{1}{4\pi^2iy}\int_{-M}^{M} dk\; \frac{ke^{iky}}{k^2 + \sigma^2},
\end{gather}
where ($y=|x-x'|$). In the limit $M \rightarrow \infty$ this is a simple contour integral. Closing the contour in the upper half plane, 
and enclosing the pole at $k=i\sigma$ one obtains
\begin{gather}
\frac{1}{4\pi y} e^{-\sigma y}.
\end{gather}
Moreover, one can evaluate the integral for finite $M$ using the exponential integral function (\emph{cf.}~\cite[p. 228]{Abram});
 one finds precisely this term, plus complicated terms involving $M$. But this term takes the form of a mass term: 
 $\sigma$ appears as an \emph{additional} mass scale in gluon correlation functions.
 
So now there are two mass scales, $M$ and $\sigma$, in marked contrast to what one expects. 
How do we get ourselves out of this quandary? One way is to choose $\sigma \simeq M$. Then the mass scale $\sigma$ is present 
and yet impossible to detect. The other way out is to force $\sigma$ to be zero.
 
It seems then, that if one makes a somewhat arbitrary choice of $\sigma=0$ or $M$ to fit the expected behaviour,
 then things work out satisfyingly. But these choices look distinctly \emph{less} arbitrary in the context
 of Kogan's and Kovner's model.  They consider the action (\ref{lomode}) as a statistical mechanical system with
 $SU(N)\times SU(N)$ symmetry at temperature $g^2(M)$. Since QCD is asymptotically free, large $M$ corresponds
 to low temperatures, and conversely. One expects that the system will undergo a phase transition from an ordered 
 state with spontaneous symmetry breaking to a disordered state with the full symmetry restored at some critical temperature. 
 One is then left to speculate as to the critical temperature (coupling) and the nature of the phase transition. In the
 context of mean field theory, $\sigma$ acts as a mass gap; for a second order transition it vanishes continuously
 at the critical point, and is otherwise non-zero. 
 
Kogan and Kovner argue that the system's internal energy will be minimised at, or close to, the critical point, with $\sigma$
taking its critical value, corresponding to the mass gap. We have argued that $\sigma$ can only take the values zero or $M$. 
Thus, the case $\sigma=0$ corresponds to no mass gap, and 
a second order phase transition, as predicted by the mean field theory. The case $\sigma \simeq M$ corresponds to a mass 
gap as big as the ultraviolet cut-off in the low mode model: clearly this corresponds to a strongly first order phase
transition. Intermediate behaviour (\emph{e.g.} a weakly first order transition) has observable consequences for the 
gluon correlators.

Collecting up the terms generated by (\ref{ellam}), that is (\ref{hiquad}), (\ref{hiquar}) and (\ref{loquar}), 
the chromo-electric correlator becomes, for $\sigma=0$
\begin{align} \label{chromoel}
4\langle E^{a 2}_{i} (x) E^{'a 2}_{i} (x) \rangle &= 
6(N^2-1)[G^{-1}(x-x')]^2 \nonumber \\
& -4  (N^2-1) G^{-1}(x-x') G^{-1} (x-x')|_{k>M} \nonumber \\
& + \left[ 8N^2 \int_{k<M} +(N^2-1) \int_{k>M} \right] \frac{d^3k}{(2\pi)^3} \frac{d^3k'}{(2\pi)^3} \;
G^{-1}(k) G^{-1}(k') \frac{(k.k')^2}{k^2 k^{'2}} e^{i(k+k').(x - x')}.
\end{align}

The final part of $\langle F^2(x) F^2(x') \rangle$ is (up to a contact term)
\begin{gather}
-8\langle B^2(x) E^2(x')\rangle = -8\langle \epsilon_{ijk}\epsilon_{ilm}\partial^{x}_{j} A^{a}_{k} \partial^{x}_{l} A^{a}_{m}
i\frac{\delta}{\delta A^{'a}_{i}(x)}i\frac{\delta}{\delta A^{'a}_{i}(x)}\rangle,
\end{gather}
where terms of $O(g^2)$ or with an odd number of $A$-fields have been discarded. 
Calculating as before, we find that this is a contact term.

After a good deal of effort, we have arrived at an expression for the correlator $\langle F^2(x)F^2(x') \rangle$ for $\sigma=0$,
namely the sum of (\ref{chromomag}) and (\ref{chromoel}). The explicit form of the correlator is not
 especially illuminating. What \emph{is} significant is that unless $\sigma=0$ or $M$, 
 there is an additional mass scale in the correlator.
\section{Pseudo-scalar gluon correlator} \label{sec:pseudo}
Let us proceed to calculate the scaling properties of the pseudo-scalar glueball correlator. In calculating the scalar correlator,
the $\sigma$ mass scale came from the quartic term in $\lambda_L$. Such a term is not present in the pseudo-scalar correlator, 
suggesting that only the expected mass scale $M$ will be present.
 
Discarding contact terms, the correlator can be written unambiguously as $16\langle BB'EE'\rangle$. 
Doing the functional differentiation in (\ref{corrdef}) and retaining terms even in $A$ of $O(g)$ gives
\begin{gather} \label{corrtwo}
16\langle\epsilon_{ijk}\partial^{x}_{j}A^{a}_{k}\epsilon^{'}_{ijk}\partial^{'x}_{j}A^{'a}_{k}
[\delta^{aa'}\delta_{ii'}G^{-1}(x-x') - \int d^3y'd^3z\; G^{-1}(y'-x')G^{-1}(z-x)A^{'a}_{i}(y)A^{a}_{i}(z)]\rangle.
\end{gather}
The first term contains two $\epsilon\partial$s only, so its only non-zero part has $\langle AA \rangle \rightarrow \mathcal{M}^{-1}$,
\emph{i.e.}
\begin{gather}
16 (N^2-1) G^{-1}(x-x') \partial^{x}_{j}\partial^{x'}_{j}G(x-x'),
\end{gather}
with the expected mass scale $M$.
The second term is more complicated. 
The possible non-zero parts in the four-point $A$ correlator in (\ref{corrtwo}) are 
\begin{align} \label{Afour}
\langle \partial^{x}_{j}A^{a}_{k}\partial^{'x}_{j}A^{'a}_{k}A^{'a}_{i}(y)A^{a}_{i}(z)\rangle &=
\partial^{x}_{j} \partial^{'x}_{j}\mathcal{M}^{-1aa'}_{\phantom{-1}kk'}(x,x')
 \mathcal{M}^{-1a'a}_{\phantom{-1}i'i}(y',z) \nonumber \\
&+\partial^{x}_{j} \mathcal{M}^{-1aa'}_{\phantom{-1}ki'}(x,y')
\partial^{'x}_{j} \mathcal{M}^{-1a'a}_{\phantom{-1}k'i}(x',z) \nonumber \\
&+2\partial^{x}_{j} \mathcal{M}^{-1aa'}_{\phantom{-1}ki'}(x,y')
\partial^{'x}_{j} a^{'a}_{k}(x) a^{a}_{i} (z) \nonumber \\
&+\partial^{x}_{j}\partial^{'x}_{j} \mathcal{M}^{-1aa'}_{\phantom{-1}kk'}(x,x') a^{'a}_{i}(y) a^{a}_{i}(z).
\end{align}

The first term yields $-8 (N^2-1) G^{-1}(x-x') \partial^{x}_{j}\partial^{x'}_{j}G(x-x')$ 
and the second term turns out to be a contact term.

The third and fourth terms on the RHS of (\ref{Afour}) have both high and low mode contributions. 
For the low mode parts, the integrations in (\ref{corrtwo}) are trivial. they result in a factor
$\langle \lambda_L (x) \lambda_L (x') \rangle$, which is $O(g^2)$.

The only remaining contribution comes from the high modes in the last term. This is
\begin{gather}
16 (N^2-1) [\partial^{x}_{j}\partial^{x'}_{i} G(x-x') \int_{k>M} \frac{d^3k}{(2\pi)^3} \; \frac{k_ik_j}{k^2} G^{-1} (k) e^{ik.(x-x')}
-\partial^{x}_{j}\partial^{x'}_{j} G(x-x') G^{-1} (x-x')|_{k>M}].
\end{gather}
Thus, the only mass scale in the pseudo-scalar glueball correlator is $M$.
\section{Discussion}
In the above, an extensive analysis of the gluon correlators in the Kogan-Kovner ansatz has been made. 
It has been shown that the gauge and Lorentz invariant gluon correlators contain the expected
dynamically generated mass scale $M$. Moreover, it has been shown that an additional mass scale $\sigma$ is present in the 
scalar glueball, 
and should be observable, \emph{unless} it vanishes, or lies close to $M$. This constrains the phase transition in the 
$\sigma$-model part of the Kogan--Kovner ansatz to be of second order, or strongly first order, respectively. If this is the case,
then the scalar and pseudo-scalar glueball masses are degenerate.
\section{Acknowledgements}
I would like to thank I. I. Kogan and A. Kovner for their comments.
%\bibliography{refs}%

\end{document}